# $B_{12}H_n$ and $B_{12}F_n$: Planar *vs* Icosahedral Structures

N. Gonzalez Szwacki[1] and C. J. Tymczak[2*]

[1]*Institute of Theoretical Physics, Faculty of Physics, University of Warsaw, ul. Hoża 69, 00-681 Warsaw, Poland*
[2]*Department of Physics, Texas Southern University, Houston, Texas 77004, USA*

Using density functional theory (DFT) and quantum Monte Carlo (QMC) calculations we show that the $B_{12}H_n$ and $B_{12}F_n$ ($n = 0$-4) quasi-planar structures are energetically more favorable than the corresponding icosahedral clusters. Moreover, we show that the fully planar $B_{12}F_6$ cluster is more stable than the 3D counterpart. These results open up the possibility of designing larger boron based nanostructures starting from quasi-planar or fully planar building blocks.

[*]E-mail: tymczakcj@tsu.edu

## 1 Introduction

The icosahedral $B_{12}H_{12}^{2-}$ cluster is the most stable molecule among the number of polyhedral boranes synthesized so far [1]. A large-scale and efficient synthesis of fully-fluorinated boron hydrides, e.g. icosahedral $B_{12}F_{12}^{2-}$, has been also reported [2]. On the other hand, the all-boron $C_{3v}$-$B_{12}$ cluster is quasi-planar and was reported to be one of the most stable all boron clusters. It was also established by extensive computations that the quasi-planar $B_{12}$ cluster is much lower in energy than the all-boron icosahedral $B_{12}$ cluster. This was reported not only for the neutral clusters [3] but also for the charged ones [4]. It is then interesting to investigate what happens with the relative stability of the two (quasi-planar and 3D) all-boron structures upon addition of hydrogen or fluorine atoms. This is the purpose of this study.

Quasi-planar and 3D boron clusters with the number of hydrogen atoms smaller than the number of boron atoms have been studied both theoretically [5-11] and experimentally [12-14]. Ohishi *et al.* [12] reported the formation of $B_{12}H_n^+$ ($n = 0$-12) cationic clusters through ion-molecule reactions of the decaborane ions ($B_{10}H_n^+$, $n = 6$-14) with diborane molecules ($B_2H_6$) in an external quadrupole static attraction ion trap. The mass spectrum analysis reveals that among $B_{12}H_n^+$ clusters with different hydrogen content $n$ the $B_{12}H_8^+$ molecule is the main product. In the same study, using first principle calculations at the B3LYP/6-31G(d) level of theory, the authors compared the relative energies of quasi-planar and 3D $B_{12}H_n^+$ clusters with $n$ varying from 0 to 12. According to that study 2D clusters with $n = 0$-5 are energetically preferred over the 3D structures, whereas 3D clusters are energetically favored for $n \geq 6$. In a more recent combined experimental/theoretical study, Ohishi *et al.* [14] suggested that quasi-planar $B_{12}H_n^+$ with $n = 0$-3 clusters can be obtained by further removal of H atoms from the decaborane ions. This opens up the possibility of changing the structure of the $B_{12}H_n^+$ cluster by controlling the number of hydrogen atoms in the cluster.

To our knowledge, there are no previous studies on the structure and properties of quasi-planar $B_{12}F_m$ clusters. However, the structures of two polyboron fluorides, $B_8F_{12}$ and $B_{10}F_{12}$, revealing unusual open structures were recently determined [15].

## 2 Theoretical approach

The initial search for the most stable structures of the boron hydrides $B_{12}H_n$ and boron fluorides $B_{12}F_n$ was done at the B3LYP/6-31G(d) level of theory using the FreeON code [16] with no symmetry restrictions. For clusters with an even number of hydrogen or fluorine atoms (even number of electrons), the computations were performed for the singlet multiplicity only, whereas doublet and quartet multiplicities were considered for cluster with an odd number of hydrogen or fluorine atoms (odd number of electrons). In the later case structures with lower multiplicity were energetically more favorable. For the charged structures a similar analysis has been done and the ground state was found to have the lowest multiplicity.

Next, the low-lying isomers have been re-optimized using the GAMESS-US code [17] and for the resulting structures, the vibrational analysis has been done to identify true local minima. Finally, for each $n$ the lowest lying isomers of $B_{12}H_n$ and $B_{12}F_n$ have been further optimized at the B3LYP/6-311++G(d,p) level of theory. The nucleus independent chemical shift (NICS) values and magnetic susceptibility tensors were calculated using the Gaussian03 package [18]. To obtain the NICS values, we have used the GIAO (gauge-independent atomic orbital) method and the magnetic susceptibility tensors were calculated using the CSGT (continuous set of gauge transformations) method.

It is important to mention that the "bare" $B_{12}$ icosahedron undergoes distortions after structural optimization and its symmetry is $S_2$ [3], not $I_h$. However, we will refer to that structure and its derivatives as icosahedral or 3D. The quasi-planar or fully planar clusters, for a change, will be often labeled as 2D structures.

The QMC calculations are done using the QWalk [19] package in two steps. The first step consists in optimizing the trial many-body wavefunction by doing Variational Monte Carlo (VMC) calculations. The trial wave function is of Slater-Jastrow form. The Slater determinants are constructed using B3LYP orbitals, generated using the GAMESS-US code with previously optimized geometries within the B3LYP/6-311++G(d,p) level of theory. For the QMC calculations, we use Gaussian basis sets with effective core potentials [20]. In the second step, we do fixed-node DMC calculations with previously optimized trial wave functions. We use a time step of 0.005 a.u. The DMC error bars are about 0.1 eV.

## 3   Planar *vs* Icosahedral Structures

The procedure for determining the most stable isomers of $B_{12}H_n$ was very similar to that reported in Ref. [12], namely we started with optimized icosahedral and quasi-planar $B_{12}$ clusters and for a given $n$ we have calculated the total energies of all possible clusters resulting from adding hydrogen atoms to the vertices of the distorted icosahedron or to the outer boron atoms of the quasi-planar structure. 2D clusters with an even $n$ have been considered in our previous work [6] and here we have extended the investigation to an odd $n$. The energetically most favorable 2D and 3D $B_{12}H_n$ structures are shown in Fig. 1. The minimum-energy cluster structures of $B_{12}F_n$, shown in Fig. 2, have been found by replacing the hydrogen atoms of the low-lying $B_{12}H_n$ isomers by fluorine atoms. Interestingly enough, the resulting structures are similar to those found for $B_{12}H_n$. One of the small differences is that the B-F bonds are on average ~13% longer than the B-H bonds.

In Fig. 3, we have plotted the total energy difference between quasi-planar (or fully planar) and icosahedral $B_{12}X_n$ (X = H, F) clusters as a function of $n$, the number of H or F atoms in the cluster. As can be seen from the figure, the quasi-planar clusters with up to 4 hydrogen atoms are more stable than the corresponding icosahedral structures (a similar result has been recently reported [11] for $B_{12}H_n^{0/-}$ clusters). The same is true for the fully planar $B_{12}F_6$ molecule, which is 0.63 eV lower in energy than the 3D cluster. The 2D and 3D $B_{12}F_5$ isomers are almost degenerated in energy. From Fig. 3 we can also see that the energy difference, $E_{2D}$-$E_{3D}$, increases monotonically with $n$, with the exception of the two "minima" for $B_{12}F_4$ and $B_{12}F_6$. These two minima may suggest an additional stabilization of the 2D structures over the 3D counterparts due to the presence of aromatic stabilization energy.

Similar results to those presented in Fig. 3 were reported for the icosahedral and quasi-planar $B_{12}H_n^+$ structures [12]. However, in their recent work, Ohishi *et al.* [14] have used the PBE0 functional instead of the B3LYP functional to determine the energies of the $B_{12}H_n^+$ clusters. The authors' choice was motivated by the fact that the B3LYP functional may overestimate the energy difference between 2D and 3D structures. To address this problem, we calculated the energy difference between the 2D and 3D structures of $B_{12}$, $B_{12}H_6$, and $B_{12}F_6$ using the very accurate DMC approach. The DMC $E_{2D}$-$E_{3D}$ values are -5.13, 0.79, and -0.47 eV for $B_{12}$, $B_{12}H_6$, and $B_{12}F_6$, respectively, whereas the corresponding B3LYP values are -5.34, 0.41, and -0.63 eV, respectively (see Fig. 3). This means that the DMC values are shifted up by a value not larger than 0.4 eV with respect to the B3LYP values. This, however, does not affect the conclusions that are drawn from Fig. 3, since even if we shift up the curves by ~0.4 eV still the quasi-planar $B_{12}H_n$ and $B_{12}F_n$ ($n$ = 0-4), and the fully planar $B_{12}F_6$ clusters remain energetically favorable.

## 4 Fully planar clusters: $B_{12}H_6$ vs $B_{12}F_6$

As calculated here and also reported in Ref. [11] the fully planar $B_{12}H_6$ cluster corresponds to a local minimum of energy, whereas the $D_{3h}$-$B_{12}F_6$ structure wins the competition with other 2D and 3D isomers and corresponds to a global minimum of energy. Many properties of the $B_{12}H_6$ cluster have been previously described in Ref. [6], but for consistency purposes, we have repeated some of those calculations at the B3LYP/6-311++G(d,p) level of theory. The HOMO-LUMO (HOMO, highest occupied molecular orbital; LUMO, lowest unoccupied molecular orbital) gaps of the planar $B_{12}H_6$ and $B_{12}F_6$ structures are 3.54 and 4.39 eV, respectively, whereas the HOMO-LUMO gaps of the corresponding 3D clusters are the same and equal to 2.73 eV. The B–H and B–F interatomic distances are 1.179 and 1.326 Å in $B_{12}H_6$ and $B_{12}F_6$, respectively. For comparison, the computed B–H and B–F bond lengths in borane ($BH_3$) and boron trifluoride ($BF_3$) are 1.190 and 1.318 Å, respectively.

While both 2D structures, $B_{12}H_6$ and $B_{12}F_6$, have similar shape and size, they exhibit quite different magnetic properties that are directly related to aromaticity. First, we have computed the anisotropy of magnetic susceptibility. The values for $B_{12}H_6$ and $B_{12}F_6$ are -208.1 and -125.8 cgs-ppm, respectively. The isotropic values of the magnetic susceptibility are -91.9 and -118.2 cgs-ppm for $B_{12}H_6$ and $B_{12}F_6$, respectively. These results suggest that the induced ring current is stronger for $B_{12}H_6$ than for $B_{12}F_6$. Similarly, as reported in Ref. [6] for $B_{12}H_6$, the central part of the $B_{12}F_6$ molecule has a paratropic current flowing inside the inner $B_3$ triangle. The antiaromaticity of the inner triangle is, however, smaller for $B_{12}F_6$ than for $B_{12}H_6$ since the NICS(0) values are 3.9 and 13.3 ppm, respectively. A global aromatic current is dominant above the $B_{12}F_6$ ($B_{12}H_6$) molecule since the NICS values are negative, NICS(1) = -5.5 ppm (-3.6 ppm) and NICS(2) = -4.8 ppm (-5.0 ppm), 1 and 2 Å above and below the center of the cluster.

To examine the influence of charge on the structure and stability of the fully planar clusters, we have also studied charged 2D and 3D $B_{12}X_6$ (X = H, F) structures. The lowest energy 2D structures identified for $B_{12}H_6$ and $B_{12}F_6$ were used as initial structures for the structural optimization at a given charge state. For the 3D structures, we have made a search over all possible configurations of the hydrogen or fluorine atoms. For the lowest energy structures with an even number of electrons, a singlet multiplicity has been assumed, whereas doublet and quartet multiplicities were considered for clusters with an odd number of electrons. In the later case clusters with lower multiplicity were energetically more favorable. The structures of the 2D and 3D charged clusters are shown in Fig. 4. It has been previously reported that the fully planar $D_{3h}$-$B_{12}H_6$ cluster undergoes structural distortions if charged with one electron, although the quasi-planarity is preserved [6]. In general, all the charged 2D $B_{12}X_6$ (X = H, F) clusters are quasi-planar rather than fully planar, what can be seen in Fig. 4. In Fig. 5 we have plotted the energy difference between 2D and 3D $[B_{12}X_6]^q$ (X = H, F) structures as a function of the cluster charge state $q$. We have found that the addition of one or two electrons to fully planar $B_{12}H_6$ and $B_{12}F_6$ clusters (or the removal of one electron from them) makes those structures even less energetically favorable with respect to the corresponding 3D isomers. This is, however, less pronounced for $B_{12}H_6$ than for $B_{12}F_6$ as shown in Fig. 5. Finally, should be noted that the quasi-planar $B_{12}F_6^{2+}$ cluster is much more stable than its 3D isomer.

## 5 Conclusions

Our DFT and QMC results show that the $B_{12}H_n$ and $B_{12}F_n$ ($n$ = 0-4) quasi-planar structures are energetically more favorable than the corresponding icosahedral clusters and that the fully planar $B_{12}F_6$ cluster is more stable than the 3D counterpart. We have also shown that negative or positive charge further stabilizes 3D over the 2D $B_{12}X_6$ (X = H, F) clusters (except for $B_{12}X_6^{2+}$, where the opposite is observed). Our findings are potentially useful for designing larger boron based nanostructures starting from quasi-planar or fully planar building blocks.

**Acknowledgements**
The authors would like to acknowledge support given by the Robert A. Welch Foundation (Grant J-

**Figure Captions**

**Fig. 1** The most stable structures of 3D and 2D $B_{12}H_n$ ($n$ = 0-8) clusters. The symmetry of each cluster is given.

**Fig. 2** The most stable structures of 3D and 2D $B_{12}F_n$ ($n$ = 1-8) clusters. The symmetry of each cluster is given.

**Fig. 3** Energy difference between quasi-planar (or planar) and icosahedral $B_{12}X_n$ (X = H, F) clusters as a function of the number of H or F atoms.

**Fig. 4** The structures of charged 3D and 2D clusters of $B_{12}H_6$ and $B_{12}F_6$. The symmetry of each cluster is provided.

**Fig. 5** Energy difference between quasi-planar (or fully planar) and icosahedral $[B_{12}X_6]^q$ (X = H, F) clusters as a function of the cluster charge state $q$.

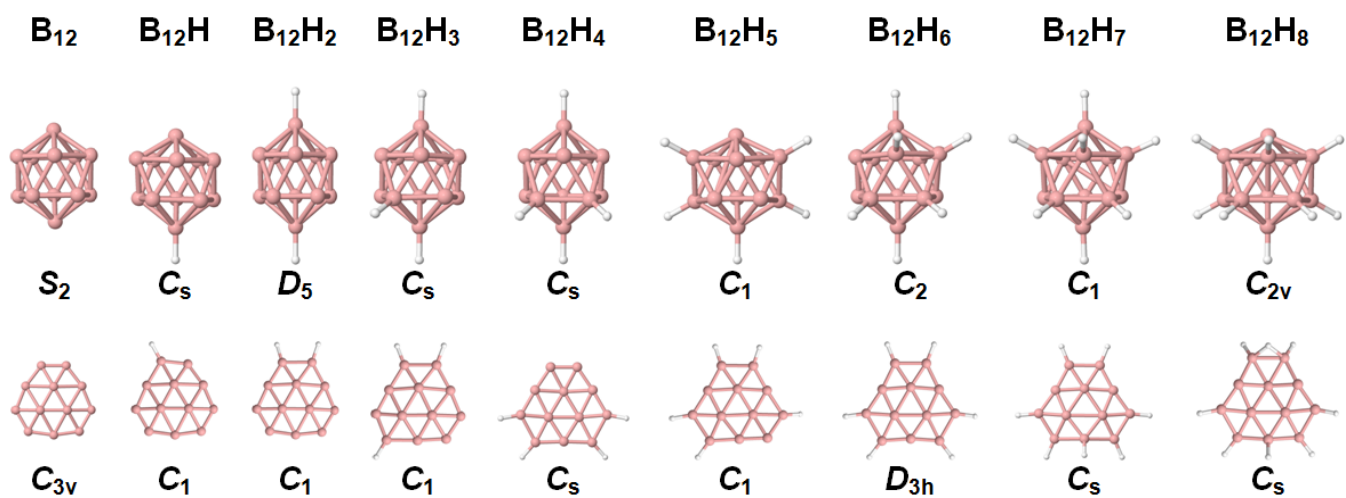

FIG. 1

**FIG. 2**

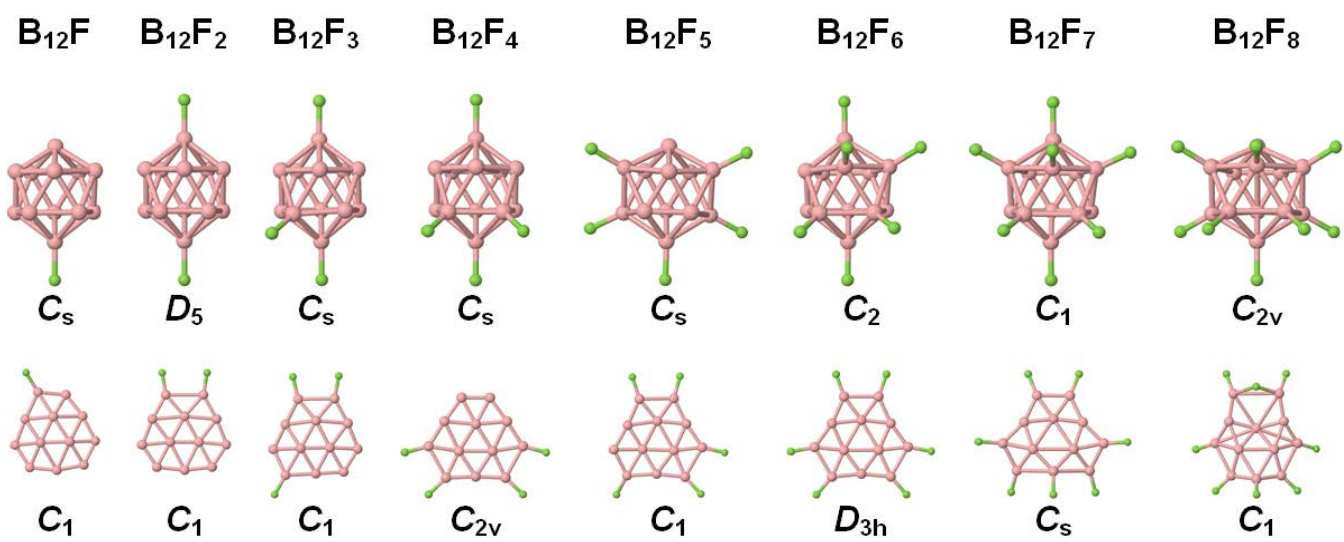

**FIG. 3**

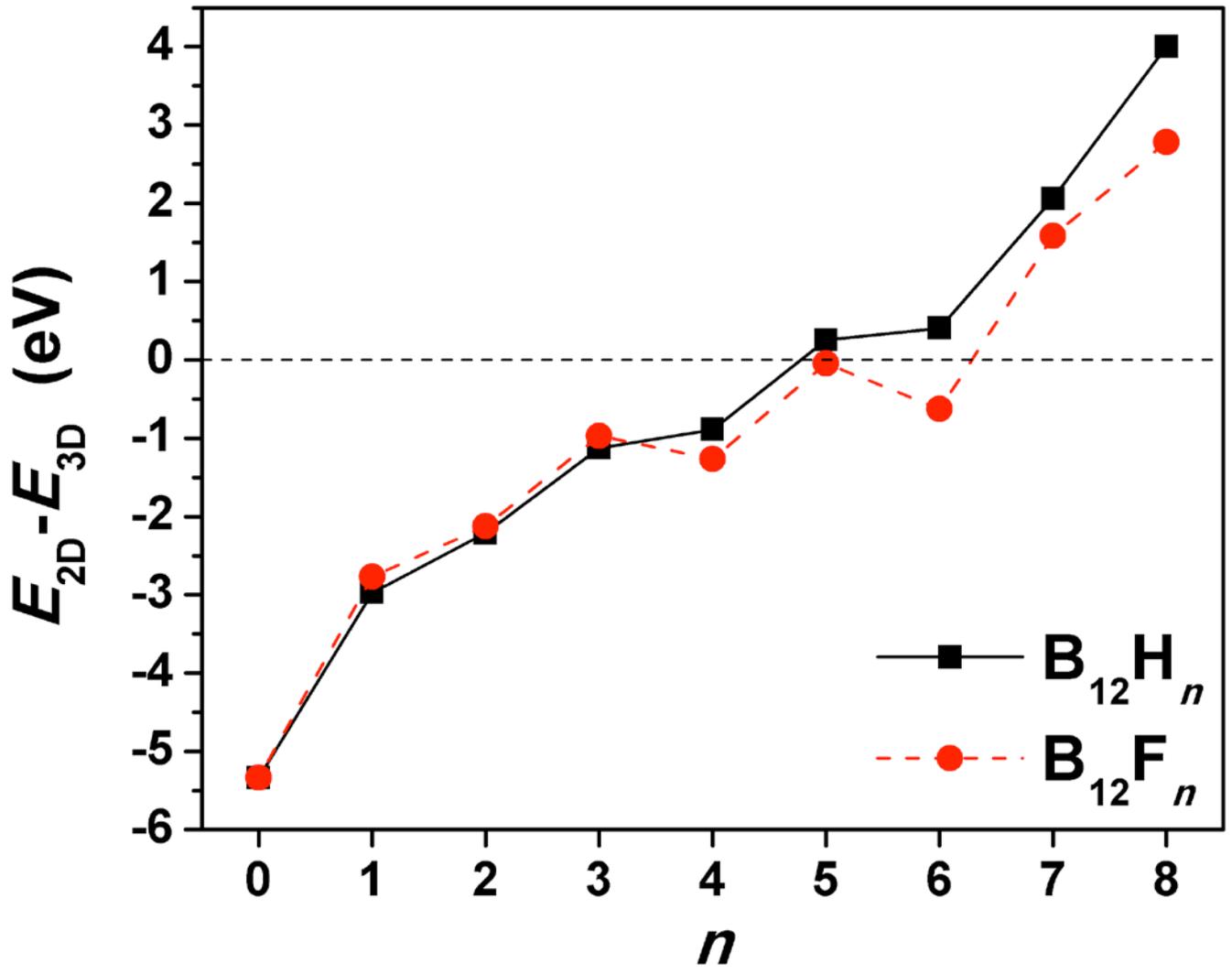

**FIG. 4**

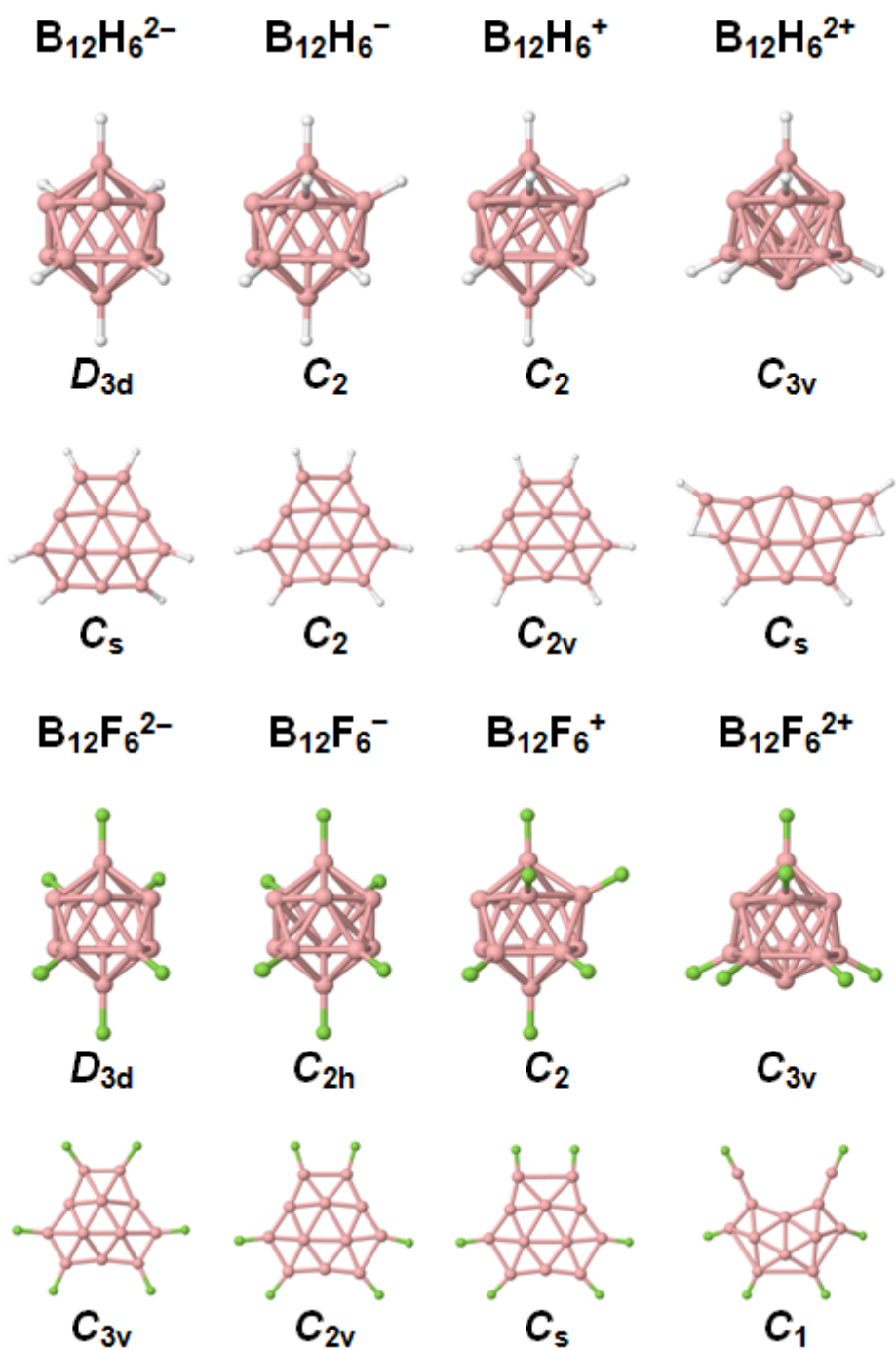

**FIG. 5**

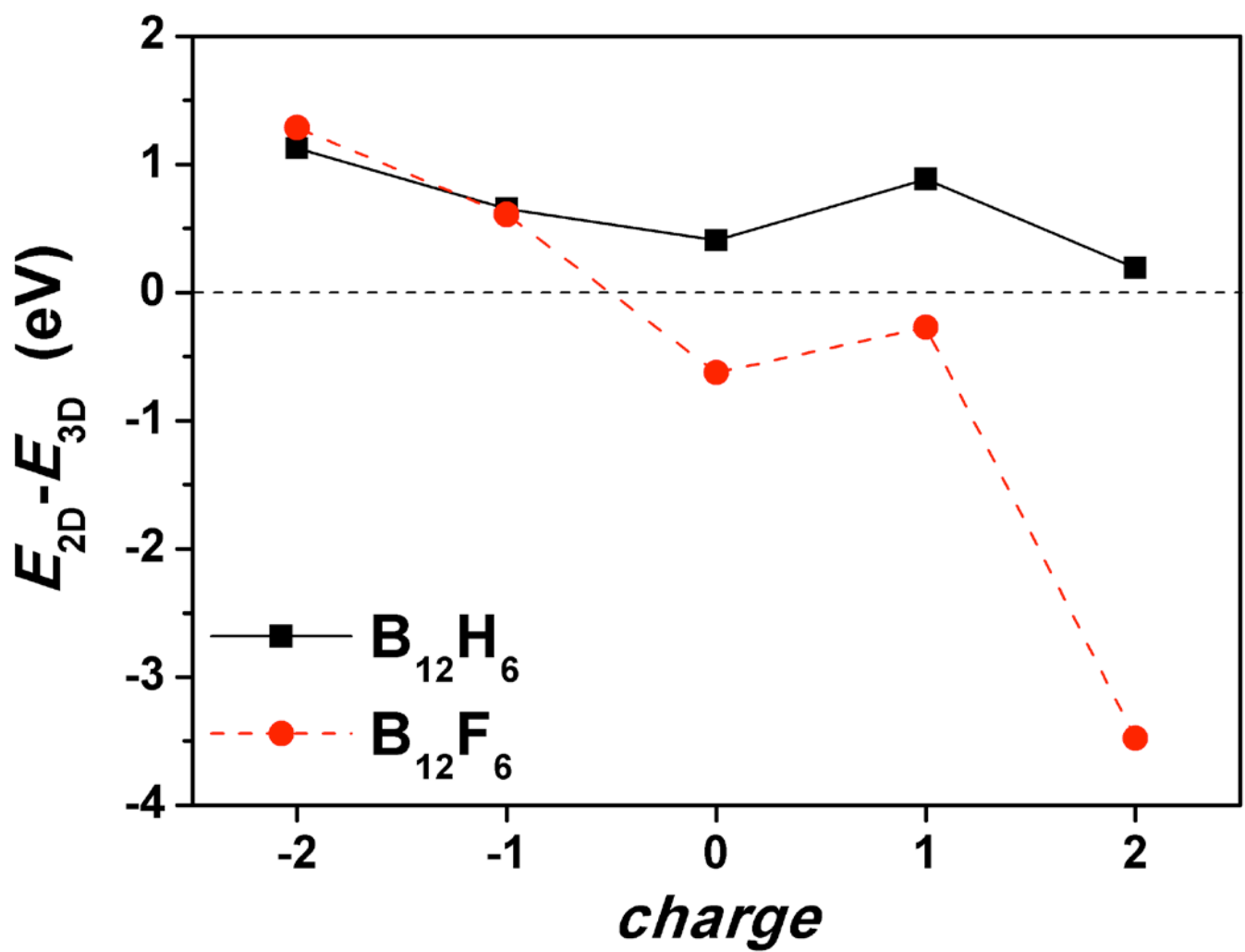